\documentclass[twocolumn,showpacs,pra,aps,superscriptaddress]{revtex4}
\usepackage{array}
\usepackage{amsmath}
\usepackage{graphicx}
\usepackage{amstext}
\usepackage{psfig}
\usepackage{amsfonts}
\begin{document}

\title{Nonclassicality of a photon-subtracted Gaussian field}

\author{M. S. Kim}

%\email{m.s.kim@qub.ac.uk}

\affiliation{School of Mathematics and Physics, The Queen's
  University, Belfast, BT7 1NN, United Kingdom}

\author{E. Park}

\affiliation{School of Mathematics and Physics, The Queen's
  University, Belfast, BT7 1NN, United Kingdom}

\author{P. L. Knight}

\affiliation{Blackett Laboratory, Imperial College, London SW7 2BW,
United Kingdom}

\author{H. Jeong}

\affiliation{Department of Physics, University of Queensland, St
Lucia, Qld 4072, Australia}

\date{\today}

\begin{abstract}
We investigate the nonclassicality of 
a photon-subtracted Gaussian field, which was produced in a recent
experiment, using negativity of the Wigner function 
and the non-existence of well-behaved positive $P$ function.  We obtain the
condition to see negativity of the Wigner function for the case
including the mixed Gaussian incoming field, the threshold photodetection and
the inefficient homodyne measurement.  We show how similar the
photon-subtracted state is to a superposition of coherent states. 

\end{abstract}


\maketitle
\section{Introduction}
The recent development of quantum optics has opened the possibility 
to generate and manipulate various non-classical light fields,
which cannot be described by classical theory, in a real laboratory.
It is generally accepted that the presence of a positive
well-defined $P$ function (a quasiprobability function in phase space
\cite{Sudarshan})
signals the field classical \cite{Mandel}; otherwise the field is
categorized as nonclassical. A stronger constraint on nonclassicality
is the presence of negativity in the Wigner function (another
quasiprobability function) of the field \cite{Lee}.  While a Gaussian field 
may not have its $P$ function, its Wigner function never
becomes negative. For example, the squeezed vacuum state is represented by its
Gaussian Wigner function while its $P$ function does not exist
\cite{Barnett}. 
It is also known that a Gaussian field remains Gaussian
by linear transformations which correspond to 
basic tools in a quantum optics laboratory
such as a phase shifter, a beam splitter and a squeezer \cite{Simon,Ekert}.  

Two better-known nonclassical fields are a squeezed state and a
superposition of two separate coherent states (coherent-state superposition).
The two kinds of states are closely related to probably the most 
fundamental and intriguing paradoxes in quantum theory, i.e.,
the Einstein-Podolsky-Rosen paradox \cite{EPR} for a two-mode squeezed
state and the Schr\"odinger's cat paradox \cite{Schr} for a
coherent-state superposition.
They are also known as useful resources
for various schemes in quantum information processing. 
A squeezed state and a coherent-state superposition manifest
different types of nonclassicality. Whereas a squeezed state is a
Gaussian field, a coherent-state superposition 
is non-Gaussian and shows 
a large amount of negativity in its Wigner function. There was an early attempt
to relate the two states through quantum noise of arbitrary
strength \cite{Carmichael}. Dakna {\em et al.} \cite{Dakna} 
considered a connection between the two states by subtracting a
precise number of photons from a squeezed field. They also showed
that any quantum state can be generated from the vacuum by application
of the coherent displacement operator and adding photons
\cite{Dakna2}.  On the other hand, it has been reported that by
squeezing a single photon state one  
can generate a state which has almost unit fidelity to
a coherent-state superposition of small amplitude \cite{Lund}.

It is only very recently that a traveling non-Gaussian field
was experimentally generated by subtracting a photon from a squeezed vacuum
by Wenger {\it et al.} \cite{Grangier}. They used a beam splitter and a
threshold detector %, which discerns there being photons and no photons,
to subtract a photon from the squeezed field, but 
the reconstructed Wigner function failed to show a negative value
\cite{Grangier}.  
It is, thus, timely to analyze the generation of a non-Gaussian
state in relation to the status of experiments.  In particular, as
such the state forms a starting point for distillation of a
continuous-variable field for quantum information processing
\cite{Eisert} and may improve the efficiency of quantum teleportation
\cite{Welsch}, the study will be of use.  In this paper, we assess the nonclassicality of a 
photon-subtracted Gaussian field and study how similar this state is
to a coherent-state superposition.  We assess negativity of the Wigner
function in conjunction with the non-existence of the positive $P$ function. 

\section{Field generated by subtracting a photon}

We would like to consider what kind of state one produces by
eliminating one photon from a simple Gaussian function.  A
single-mode Gaussian field of its density operator $\hat{\rho}$ may be
represented by the Weyl characteristic function \cite{note1}
defined as
$C(\xi)=\mbox{Tr}[\hat{D}(\xi)\hat{\rho}]$: 
\begin{equation}
C(\xi)=\exp\left(-{A\over 2}\xi_r^2-{B\over 2}\xi_i^2 \right),
\label{1}
\end{equation}
where $A$ and $B$ are determined by the quadrature variances of the
field.  The displacement operator has been defined as
$\hat{D}(\xi)=\exp(\xi\hat{a}^\dag-\xi^*\hat{a})$, where
$\hat{a}$ and $\hat{a}^\dag$ are bosonic
annihilation and creation operators respectively.
Note also that the density operator can be obtained from the characteristic function as
\begin{eqnarray}
\hat{\rho}
={1\over\pi}\int d^2\xi C(\xi)\hat{D}(-\xi),
\label{character-single}
\end{eqnarray}
which can be straightforwardly obtained using
identities $(1/\pi)\int
d^2\alpha|\alpha\rangle\langle\alpha|=\openone$ and \cite{Cahill} 
$$
|\alpha\rangle\langle\beta|={1\over\pi}\int d^2\xi \hat{D}(-\xi)
\langle\beta|\hat{D}(\xi)|\alpha\rangle
$$
where $|\alpha\rangle$ is a coherent state of amplitude $\alpha$. Even
though Eq.~(\ref{1}) does not  
represent a very general Gaussian field, rotation and/or
displacement operation brings any Gaussian field to this form.   It is
useful to start with (\ref{1}) because it is extremely challenging to
produce a pure squeezed state with $AB=1$ and the characteristic function
(\ref{1}) allows us to treat a single-mode Gaussian state of a mixed state.
The uncertainty relation is given by $AB\geq 1$ and the Gaussian
state is called squeezed when either $A<1$ or $B<1$.    

Let us consider the experiment by Wenger {\em et al.} \cite{Grangier}.
First of all, they produce a squeezed Gaussian state then this
passes through a beam splitter with its transmittivity $T=t^2$,
where the other input port is assumed to be served by a vacuum.
At the one output of mode 2, we conditionally measure a one
photon state $|1\rangle_2$.  The state obtained at the other
output port of mode 1 was what Wenger {\em et
al.}  produced as a non-Gaussian field in their experiment. %Let us
We will evaluate the Wigner function for this field of mode 1.

By beam splitting the squeezed Gaussian field whose characteristic
function is written as  (\ref{1}) and
the vacuum of its characteristic function, $C_v(\xi)=\exp(-{1\over 2}
|\xi|^2)$,  the characteristic function for the output field of
modes 1 and 2 is \cite{KimLee}
\begin{equation}
C_{out}(\eta,\xi)=\exp\left(-{1\over 2}{\bf x V}{\bf x}^T\right)
\label{character-out}
\end{equation}
where ${\bf x} = (\eta_r, \eta_i, \xi_r, \xi_i)$ and the
correlation matrix
\begin{equation}
{\bf V}=
\begin{pmatrix}
n_1 & 0 & c_1 & 0 \\
0 & n_2 & 0 & c_2 \\
c_1 & 0 & m_1 & 0 \\
0 & c_2 & 0 & m_2
\end{pmatrix}
\label{correlation}
\end{equation}
with 
\begin{eqnarray}
n_1=TA+R,~~n_2=TB+R,~~c_1=tr(A-1),
\nonumber \\
c_2=tr(B-1),~~ m_1=RA+T,~~m_2=RB+T
\label{value}
\end{eqnarray}
and $T=t^2$ and $R=r^2$. 

We then use the two-mode
version of (\ref{character-single}) for the density operator of the
output field:
\begin{equation}
\hat{\rho}_{out}=\frac{1}{\pi^2}\int
C_{out}(\eta,\xi)\hat{D}_1(-\eta) \hat{D}_2(-\xi)d^2\eta d^2\xi.
\label{rho-out}
\end{equation}
The density operator for the field of mode 1 conditioned on
one-photon measurement for mode 2 is
\begin{equation}
\hat{\rho}_1={\cal N}~_2\langle 1|\hat{\rho}_{out}|1\rangle_2.
\label{rho-1}
\end{equation}
Throughout the paper, ${\cal N}$ denotes the appropriate normalization
factor.  For the case of Eq.~(\ref{rho-1}), 
\begin{equation}
{\cal N}=\frac{1}{_2\langle
  1|\mbox{Tr}_1[\hat{\rho}_{out}]|1\rangle_2}=
  \frac{[(m_1+1)(m_2+1)]^{3/2}}{2(m_1m_2-1)}.
\label{normal}
\end{equation}
With the knowledge of the one-photon Fock state expectation value of
the displacement operator \cite{Cahill,KimKnight}
$$
\langle 1|\hat{D}(-\xi)| 1\rangle=\mbox{e}^{-{|\xi|^2\over 2}}
(1-|\xi|^2),
$$
the density operator is found to be
$$
\hat{\rho}_1={{\cal N}\over\pi^2}\int C(\eta,\xi)\hat{D}_1(-\eta)
\mbox{e}^{-{|\xi|^2\over 2}}(1-|\xi|^2)d^2\eta d^2\xi.
$$
The characteristic function is then easily obtained using the
identity
$\mbox{Tr}[\hat{D}(\zeta)\hat{D}(-\eta)]=\pi\delta^{(2)}(\zeta-\eta)$:
\begin{widetext}
\begin{equation}
C_1(\zeta) = \Big[1- \frac{c_1^2(m_2+1)\zeta_r^2}{(m_1+1)(m_1m_2-1)}
- \frac{c_2^2(m_1+1)\zeta_i^2}{(m_2+1)(m_1m_2-1)} \Big]
\exp\Big[-{1\over 2}\Big(n_1-{c_1^2\over m_1+1}
\Big)\zeta_r^2 - {1\over 2}
  \Big(n_2-{c_2^2 \over m_2+1} \Big)\zeta_i^2\Big].%\nonumber \\
\label{character-1}
\end{equation}
\end{widetext}
By Fourier transformation of the Weyl characteristic function \cite{Cahill2},
we obtain the Wigner function.
Now, the first point we are interested
in is the negativity of the Wigner function.
It is clear that the Fourier transform of (\ref{character-1}) has the
largest negativity (if any exists) at the origin of phase space and the value
of the Wigner function at the point is
\begin{equation}
W_1(0)\propto \frac{B-1}{(T+1)B+R}+\frac{A-1}{(T+1)A+R}
\label{W-0}
\end{equation}
which has been obtained by substituting the parameters (\ref{value}).
It is obvious that if $A>1$ or $B>1$, {\em i.e.}, the 
incoming Gaussian
field is not squeezed, $W(0)$ is positive everywhere.  In order to
find the exact condition for negativity in the Wigner function, we
assume that $A<1, ~B>1$ and introduce positive parameters
$x=(A+1)/(1-A)$ and $y=(B+1)/(B-1)$.  Then the right-hand-side (RHS) of
(\ref{W-0}) becomes
$$
\frac{2T-x+y}{(T-x)(T+y)}
$$
whose denominator is always negative.  The numerator becomes positive
when the transmittivity satisfies 
\begin{equation}
T>\frac{AB-1}{(1-A)(B-1)}
\label{condition-W1}
\end{equation}
which always holds when the incoming Gaussian field is pure $AB=1$ (in
other words, if the incoming Gaussian field is a pure squeezed state,
the Wigner function always shows negativity by subtracting a photon
from it).  

The $P$ function of the field may be obtained using the relation
between its characteristic function $C^{(p)}_1$ and the Weyl
characteristic function \cite{Cahill2}:
\begin{equation}
C^{(p)}(\zeta)=C(\zeta)\mbox{e}^{\frac{1}{2} |\zeta|^2}.
\label{relations}
\end{equation}
With use of the characteristic function (\ref{character-1}) and
general Gaussian integration, we find that the $P$
characteristic function is integrable when $(n_i-1)(m_i+1)-c_i^2>0$
for $i=1,2$.  By substituting the parameters (\ref{value}), we
find the condition equivalent to $2T(A-1)>0$ and $2T(B-1)>0$.  So if
the incoming field is squeezed, it is not possible to integrate
$C^{(p)}$ and no $P$ function exists.  Considering the positivity of
the $P$ function, after a little algebra with Fourier transformation of
the $P$  characteristic function, we find that the $P$ function is positive
as far as it exists in this case.  We conclude that the single-photon
subtracted field is nonclassical (in the sense of a lack of an
acceptable $P$ function) provided the original incoming field is
squeezed. (However, the Wigner function does not necessarily show
negativity for all those 
nonclassical states unless the incoming Gaussian field was pure.)
Unless the incoming Gaussian field is nonclassical we cannot generate
a nonclassical state by subtracting a photon from it.  

This seemingly trivial result %shown above
is not obvious at all as contrasted by the
nonclassicality of a field by adding a photon into a Gaussian field
\cite{Agarwal,Lee0}.
In distinction to the case of
subtracting a photon, the photon-added Gaussian state always shows
negativity at the origin of the phase space
\cite{Lee0,Mandel2,Lee1}.
By adding a photon, a highly classical state
such as a high-temperature thermal state
becomes non-classical, showing negativity in its Wigner
function.  The realization of such a photon added
state is beyond the scope of the current work but we may think of a
possibility within cavity quantum electrodynamics or the phonon state
of a driven ion in a cavity \cite{Lee1}.
 
We now introduce the coherent-superposition state
\cite{Knight} 
\begin{equation}
\label{cat-state}
|\psi\rangle = {\cal N}(|\alpha\rangle-|
-\alpha\rangle),
\end{equation}
where ${\cal N}=1/\sqrt{1-\exp[-2\alpha^2]}$, 
to assess its fidelity to the photon-subtracted Gaussian state.
It is straightforward to calculate the characteristic function
of the coherent-state superposition from
Eq.~(\ref{cat-state}) \cite{Barnett}.
The closeness of two states, one of which is a pure state
$|\phi\rangle$ and the other (pure or mixed) is represented by its
density operator $\hat{\rho}$, is measured by fidelity $\cal F$:
\begin{equation}
{\cal F} = \langle\phi|\hat{\rho}|\phi\rangle %\nonumber\\
={1\over\pi}\int d^2\zeta C_{\phi}(\zeta)C_\rho(\zeta)
\label{fidelity}
\end{equation}
where the subscripts refer to the respective states. 

\begin{figure}
\begin{center}
\includegraphics[width=0.4\textwidth]{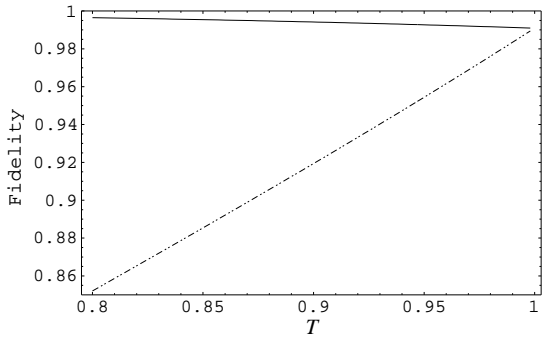}
\caption{The fidelity between 
 the photon-subtracted state 
 and the ideal coherent-state superposition 
 with an ideal single-photon detector (solid line)
  and a threshold detector (dotted line).  
The initial squeezing  parameter is $\exp(2s)=2.36$
and the $x$ axis is the transmittivity of the beam splitter $T=t^2$.
The amplitude $\alpha$ of the ideal coherent-state superposition is optimized
for the maximum fidelity.
 The optimized amplitude $\alpha$ ranges between  
 $1.02~({\rm when~} T=0.8)$ and $1.16~({\rm when~} T\rightarrow1)$.}
  \label{fig:entangle2}
  \end{center}
\end{figure}

The fidelity between
$\hat{\rho}_1$ and the coherent-state superposition (\ref{cat-state})
has been calculated from Eqs.~(\ref{1}), %(\ref{cc})
(\ref{cat-state}), and (\ref{fidelity})
and plotted in Fig.~\ref{fig:entangle2}.
The incoming Gaussian field 
has been assumed a pure squeezed field.
In Fig.~\ref{fig:entangle2}, the solid line is the 
optimized fidelity
between the photon-subtracted state and the ideal
coherent-state superposition
by an ideal single-photon detector. 
The fidelity is very high as ${\cal F}>0.99$ %which is evident in the figure.
regardless of the transmittivity of the beam splitter when
an ideal single-photon detector is used. 
The optimized amplitude of the ideal coherent-state superposition
is $\alpha=1.16$ for the transmittivity
close to unity. 
If the transmittivity $T$ gets smaller, the 
amplitude of the ideal coherent-state superposition,
which maximizes the fidelity, also becomes smaller.
For example, the amplitude will be $\alpha=1.02 (1.09)$ for $T=0.8$ (0.9).
However, the fidelity is not sensitive to the transmittivity of
the beam splitter as shown in Fig.~\ref{fig:entangle2}
because the single photon detector 
successfully subtracts only one photon from the Gaussian state
regardless of the transmittivity of the beam splitter.
In fact, the fidelity gets slightly better as the transmittivity
becomes smaller, due to the fact that both of the states
are reduced to the exact single photon state as $T\rightarrow0$.

It is interesting that the fidelity between the 
photon subtracted field $\hat{\rho}_1$ and the
coherent-state superposition is very high. This could
have been guessed from their photon number distributions.  The
squeezed vacuum is a state with only an 
even number of photons \cite{Barnett} while the coherent-state superposition
(\ref{cat-state}) is a state with only an odd number of photons
\cite{Lund}.  By subtracting one photon from the squeezed state, the
two states may become closer to each other.  We see that the
photon-subtracted squeezed field is close to the coherent-state
superposition of small amplitudes.  One reason can be found again in
their photon number distributions.  The photon number distribution of
$|\psi\rangle$ peaks around $|\alpha|^2$ while that of $\hat{\rho}_1$
is a monotonous decreasing function with regard to the photon number.
Thus, when $\alpha$ is small, the distributions become similar to each
other.  Of course, this 
check of the photon number distributions gives only a hint as the
photon number distribution does not necessarily convey all the
coherence properties of a quantum field.

\section{Experimental Reality}
As can be seen in Fig. 1, the fidelity between the ideal
coherent-state superposition and the photon-subtracted state is not so
sensitive to reflectivity of the beam splitter.  
This seemingly good result is due to an ideal
single photon detector assumed for the photon-subtracted
state $\hat{\rho}_1$. As mentioned, the state
(\ref{rho-1}) is what 
is wanted to achieve but the available high-efficiency photodetector
is not able to discern 1 and any number of photons.  Thus the state
experimentally generated using such the threshold photodetector is
\begin{equation}
\hat{\rho}_a={\cal N}\sum_{n=1}^{\infty}~_2\langle
n|\hat{\rho}_{out} |n\rangle_2 \label{any-rho}.
\end{equation}
Consider the density operator for mode 1 of the output field
\begin{equation}
\hat{\rho}_t=\mbox{Tr}_2[\hat{\rho}_{out}]=\sum_{n=0}^{\infty}
~_2\langle n|\hat{\rho}_{out}|n\rangle_2. \label{rho-t}
\end{equation}
It is then clear from Eqs.(\ref{any-rho}) and (\ref{rho-t}) that
\begin{equation}
\hat{\rho}_a={\cal N}(\hat{\rho}_t-~_2\langle 0|\hat{\rho}_{out}
|0\rangle_2) \label{new-rho}
\end{equation}
where
\begin{equation}
\hat{\rho}_t={1 \over \pi}\int C_{out}(\eta,0)
\hat{D}_1(-\eta)d^2\eta \label{101}
\end{equation}
and
\begin{equation}
_2\langle 0|\hat{\rho}_{out}|0\rangle_2= {1 \over \pi^2}\int
C_{out}(\eta, \xi)\mbox{e}^{-|\xi|^2/2}\hat{D}_1(-\eta)d^2\eta
d^2\xi. \label{102}
\end{equation}
Using $C(\eta,\xi)$ we have already discussed, we find the
characteristic function $C_a(\zeta)$ for $\hat{\rho}_a$:
\begin{eqnarray}
C_a(\zeta)&=&{\cal N}\mbox{e}^{-{1\over
2}(n_1\zeta_r^2+n_2\zeta_i^2)} \Big[1 -
\frac{2}{\sqrt{(m_1+1)(m_2+1)}}\nonumber \\
&\times&\exp\Big(\frac{c_1^2}{2(m_1+1)}\zeta_r^2
+ \frac{c_2^2}{2(m_2+1)}\zeta_i^2\Big)\Big]. \label{103}
\end{eqnarray}
The normalization factor is calculated as
$$
{\cal N}=\frac{\sqrt{(m_1+1)(m_2+1)}}{\sqrt{(m_1+1)(m_2+1)}-2}.
$$
The Wigner function obtained by Fourier transformation of the
characteristic function (\ref{103}) is what Wenger {\em et al.} would have
reconstructed \cite{Grangier} if the detection efficiency of their
experiment had been perfect and the modal purity unity.   

Let us next consider the negativity of the Wigner function.  By
inspection of the characteristic function, we realize that the Wigner
function has the largest negativity (if any) at the origin of the
phase space and the value of the Wigner function at this point is
\begin{eqnarray}
&&
W_a(0)=\frac{2{\cal N}}{\pi}\Big\{\frac{1}{\sqrt{n_1n_2}}
\nonumber \\
&&~~~
- \frac{2}{\sqrt{[n_1(m_1+1)-c_1^2][n_2(m_2+1)-c_2^2]}} \Big\}
\label{Wigner-a-origin}.
\end{eqnarray}
By partly substituting the parameters (\ref{value}), we find that the
Wigner function becomes negative when
\begin{equation}
\frac{2}{\sqrt{(n_1+A)(n_2+B)}}>\frac{1}{\sqrt{n_1n_2}}
\label{Wigner-a-condition}
\end{equation}
which becomes a criterion for the transmittivity
\begin{equation}
T>\frac{4-(A+1)(B+1)}{3(A-1)(B-1)}.
\label{Wigner-a-condition2}
\end{equation}
For a pure squeezed Gaussian incoming field, the condition becomes
$T>1/3$.  It is interesting to note that regardless of the degree of
squeezing (provided it is not zero), we can see
the negativity in the Wigner function provided the transmittivity is
larger than 1/3.

Let us assess the degree of nonclassicality by the $P$ function
criterion.  With 
use of the relation (\ref{relations}) between the characteristic
functions, we note that the $P$ characteristic function for 
$\hat{\rho}_a$ is integrable when $T(A-1)>0$ and $T(B-1)>0$. We have
checked that the $P$ function is semi-positive when it exists and
conclude that, for
nonzero transmittivity of the beam splitter, iff the incoming Gaussian
field is squeezed, the any-number photon subtracted state
$\hat{\rho}_a$ is nonclassical.  Again, the $P$ function criterion is
weaker than the negativity criterion for the Wigner function.

We now consider how close the field obtained using the threshold
detector to the coherent-state superposition (\ref{cat-state}).  The
optimized fidelity has been calculated 
using Eqs.~(\ref{cat-state}, \ref{fidelity}) and (\ref{103}),
and plotted in Fig.\ref{fig:entangle2}. It tells us that the state 
which is obtained by subtracting any number of photons is similar
to the %cat
coherent-state superposition only when the transmittivity of the
beam splitter is very high. For example,
the fidelity is higher than 90\% when $T>0.87$.
 In this case, the chance of one
photon subtraction is more likely.
Note that the optimized amplitude $\alpha$ ranges between  
 $1.02~({\rm when~} T=0.8)$ and $1.16~({\rm when~} T\rightarrow1)$
 in Fig.~\ref{fig:entangle2}.

\begin{figure}
\centerline{(a)}
\centerline{\scalebox{0.7}{\includegraphics{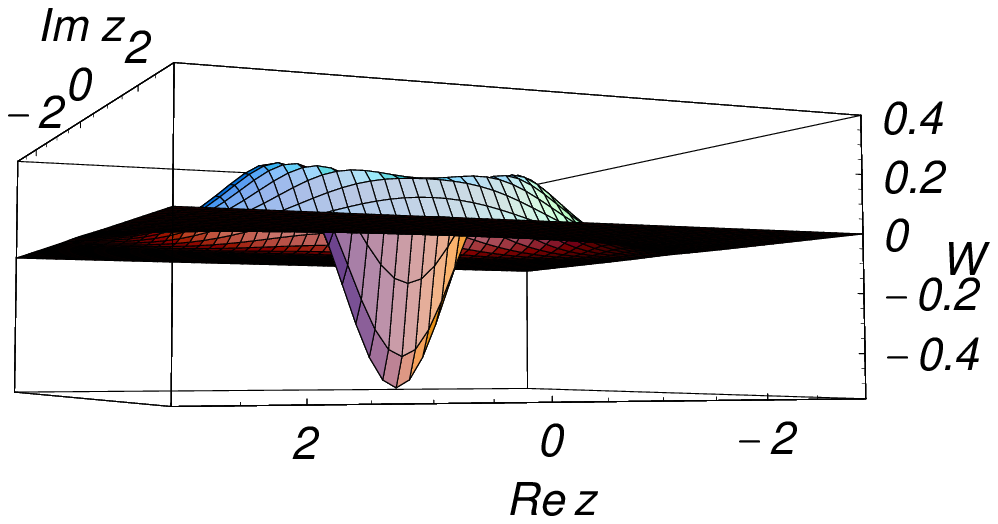}}}
\vspace{0.2cm}
\centerline{(b)}
\centerline{\scalebox{0.7}{\includegraphics{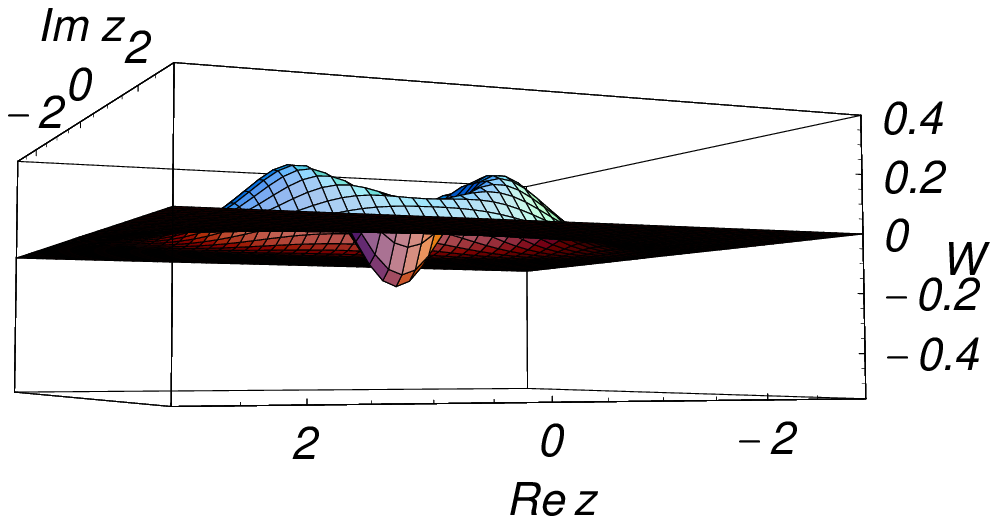}}}
\vspace{0.2cm}
\centerline{(c)}
\centerline{\scalebox{0.7}{\includegraphics{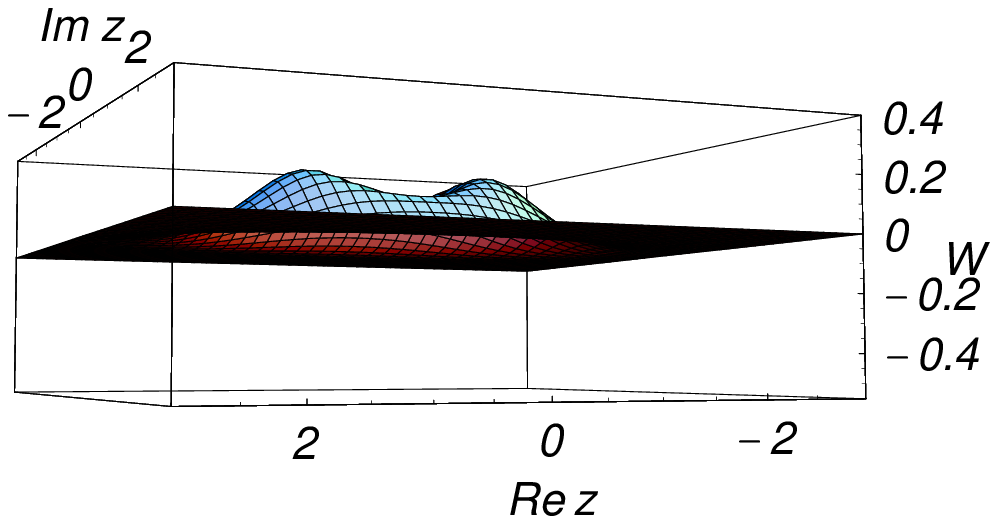}}}
\caption{(Color online) (a) The Wigner function $W(z)$ of
a photon-subtracted Gaussian state
with a threshold detector for photon subtraction and
ideal homodyne detectors for reconstruction of the Wigner function
when  %transmittivity of the beam splitter 
$T=0.88$ and $\exp(2s)=2.36$.
The minimum negativity is found as $W(0,0)=-0.52$. 
(b) The Wigner function of a photon subtracted Gaussian state
under the same condition as (a) but with homodyne efficiency
$\eta=0.75$. The minimum negativity has been reduced to $-0.15$.
(c) The Wigner function of a photon subtracted Gaussian state
under the same condition as (a) but with homodyne efficiency
$\eta=0.75$ and with the modal purity factor $0.7$. The negativity
of the Winger function has disappeared as $W(0)=0.075$.
}
  \label{fig:entangle3}
\end{figure}

\subsection{Inefficient detection and modal purity}
Homodyne detection may be used to reconstruct the Wigner function for the field
$\hat{\rho}_a$.  Even though homodyne detectors are known for their high
efficiency, the overall detection efficiency was about 75\% in Wenger
{\em et al}.'s experiment \cite{Grangier}.  An
imperfect detector is equivalent to a perfect detector with a beam
splitter in front \cite{Yuen}, where the transmittivity of the beam splitter is
determined by the detection efficiency $\eta$.  From ref.
\cite{KimImoto}, we note that the characteristic function for the
signal field passing through a beam splitter where the other input
port is served by the vacuum is
\begin{equation}
C_{im}(\zeta)=C_a(\sqrt{\eta}\zeta)C_v(\sqrt{1-\eta}\zeta).
\label{imperfect-char}
\end{equation}
Substituting Eq.~(\ref{103}) into Eq.~(\ref{imperfect-char}), we find
the characteristic function for the detected field.  The Fourier
transform of the characteristic function shows its largest negativity
at the origin of the phase space and the value there is
\begin{equation}
W_{im}(0)\propto \frac{1}{\sqrt{vw}}-\frac{1}{\sqrt{
    \Big(v-\frac{R(A-1)}{2}\Big)\Big(w-\frac{R(B-1)}{2}\Big)}}
\label{imperfect-Wigner-0}
\end{equation}
where $v=T(A-1)\eta+1$ and $w=T(B-1)\eta+1$.  Under the assumption
$(A-1)(B-1)<0$, this becomes negative when the detection efficiency satisfies
$$
\eta>-\frac{1}{2T(A-1)}-\frac{1}{2T(B-1)}-\frac{R}{4T}.
$$
In particular, for a pure Gaussian incoming field, the condition
becomes
\begin{equation}
\eta>\frac{1+T}{4T}.
\label{condition-imperfect}
\end{equation}
The RHS is smaller than unity (the detection efficiency $\eta\leq 1$)
only when $T\geq 1/3$.  This is in good agreement with the perfect
detection case.  So, in order to see negativity in the Wigner
function, the beam splitter has to have a transmittivity
larger than 1/3 first and then the detection efficiency has to satisfy
the condition (\ref{condition-imperfect}).  Wenger {\em et al.}
employed a beam splitter with $T\approx 0.88$ in which case the detection
has to be larger than a mere 53.4\% to see negativity in the
Wigner function.

Another important factor which degrades the quantum effect of the
photon subtracted 
Gaussian state in a real experiment is the modal purity factor
\cite{Grangier}.  
If the dark count rate of the photodetector employed to subtract a photon 
is non-negligible, the resulting state can be estimated in a mixture of photon
subtracted squeezed state and squeezed state as
\begin{equation}
\xi W(\alpha)+(1-\xi)W_{sq}(\alpha)
\end{equation}
where $W(\alpha)$ is the Wigner function of the photon subtracted
squeezed state, 
$W_{sq}(\alpha)$ is the Wigner function of the squeezed state, and
$\xi$ corresponds to the modal purity factor, which was 0.7,
in Wenger {\it et al.}'s experiment \cite{Grangier}. 
The Wigner functions of the photon-subtracted Gaussian state 
have been plotted for a number of different cases in Fig.~\ref{fig:entangle3}.
It shows that the negativity of the Wigner function disappears when
both of the homodyne efficiency $\eta$ and the modal purity $\xi$
are considered taking relevant experimental values. 
We suggest that either the homodyne efficiency should be improved from
0.75 to 0.9
or the modal purity factor should be improved from 0.7 to 0.9 
to clearly observe the negativity of the Wigner function.
In these cases, the minimum negativity will be $-0.044$ and
$-0.073$, respectively.

\section{Remarks}  
In this paper, we are interested in the nonclassicality of a state
produced by subtracting photons from a Gaussian field.  Subtracting a
photon does not transform a classical state into a nonclassical state
whereas a nonclassical input remains nonclassical.  This is
in contrast to the case of adding a photon to a Gaussian 
field, in which case even a very chaotic field transforms into a
nonclassical state \cite{Lee0,Mandel2,Lee1}. 
The non-Gaussian state obtained by subtracting a
photon from a Gaussian field may show large negativity in its Wigner
function.  The condition to obtain the negativity is analyzed for a
realistic case including the mixed-state input, threshold detection,
inefficient homodyne detection, and modal purity.  The non-Gaussian state analyzed in
this paper is compared with a coherent-state superposition which
may be extremely useful for fundamental and application reasons.
The comparison
shows the fidelity higher than 90\% for the experimentally relevant situation. 
We compare our analysis with a recent experimental demonstration
 \cite{Grangier} of the photon-subtracted Gaussian field
and suggest that either the homodyne efficiency 
or the modal purity factor should be improved to around $\sim 0.9$ 
to clearly observe the negativity of the Wigner function.

{\em Note added} After completion of the work, we were made aware of
ref.~\cite{nonlocality} which considers nonlocality of a photon
subtracted squeezed state.

\acknowledgements 
This work was supported by the UK Engineering
and Physical Science Research Council, the European Union and the
Korea Research Foundation (2003-070-C00024). H.J. acknowledges
the Australian Research Council for financial support.


\begin{references}
\bibitem{note1} There are two
  characteristic functions in the paper. They are the Weyl and $P$
  characteristic  functions whose Fourier transformations are,
  respectively, the Wigner and $P$ functions.  Throughout the paper,
  unless precisely mentioned, a characteristic function denotes the
  Weyl characteristic function.
\bibitem{note2} For the basic technique of the transformation,
one can refer to Appendix 12 of Ref.~\cite{Barnett}.
\end{references}
\end{document}